\documentclass[a4paper,11pt]{article}
\usepackage{pos}
\usepackage{enumitem}

\title{Isospin-breaking corrections to weak decays: \\
the current status and a new infrared improvement}
\ShortTitle{Isospin-breaking corrections to weak decays}

\author*[a,b]{Matteo Di Carlo}

\affiliation[a]{Theoretical Physics Department, CERN,\\ 1211 Geneva 23, Switzerland}
\affiliation[b]{School of Physics and Astronomy, University of Edinburgh,\\ Edinburgh EH9 3FD, United Kingdom}

\emailAdd{matteo.dicarlo@cern.ch}

\abstract{We provide an overview of the current state of lattice calculations concerning isospin-breaking corrections in weak decays, focusing in particular on leptonic decays of light pseudoscalar mesons.  We examine the two currently existing calculations, placing a particular emphasis on the role of electromagnetic finite-volume corrections. Given the large systematic uncertainty associated with such corrections, we propose a novel method to improve the finite-volume scaling of leptonic decay rates and other hadronic observables, like hadron masses and the hadronic vacuum polarization contribution to the muon $g{-}2$. We introduce a new prescription for the QED action, referred to as QED$_\mathrm{r}$, which effectively removes finite-volume corrections at $\mathcal{O}(1/L^3)$ and consequently shifts the systematic uncertainty to a higher order.  The specifics of this action, the practical implementation of the infrared improvement, and the ongoing numerical tests are also discussed.}

\FullConference{The 40th International Symposium on Lattice Field Theory (Lattice 2023)\\
July 31st - August 4th, 2023\\
Fermi National Accelerator Laboratory\\}

\newcommand{\pvec}{{\mathbf{p}}}
\newcommand{\kvec}{{\mathbf{k}}}
\newcommand{\nvec}{{\mathbf{n}}}
\newcommand{\vvec}{{\mathbf{v}}}
\newcommand{\xvec}{{\mathbf{x}}}
\newcommand{\zero}{\boldsymbol{0}}

\newcommand{\aem}{\alpha_{\mathrm{em}}}
\newcommand{\plTwo}{\pi_{\ell 2}}
\newcommand{\KlTwo}{K_{\ell 2}}
\newcommand{\pmuTwo}{\pi_{\mu 2}}
\newcommand{\KmuTwo}{K_{\mu 2}}

\newcommand{\QEDTL}{\text{QED$_\text{TL}$}}
\newcommand{\QEDSF}{\text{QED$_\text{SF}$}}
\newcommand{\QEDL}{\text{QED$_\text{L}$}}
\newcommand{\QEDr}{\text{QED$_\text{r}$}}
\newcommand{\QEDC}{\text{QED$_\text{C}$}}
\newcommand{\QEDm}{\text{QED$_\text{m}$}}

\newcommand{\QCD}{\text{QCD}}

\newcommand{\Mcal}{\mathcal{M}}
\newcommand{\Ocal}{\mathcal{O}}
\newcommand{\Scal}{\mathcal{S}}

\newcommand{\dd}{\mathrm{d}}

\newcommand{\ii}{\mathrm{i}}

\newcommand{\nn}{\nonumber}

\usepackage{cleveref}
\crefformat{section}{section~#2#1#3}
\crefmultiformat{section}{sections~#2#1#3}
    { and~#2#1#3}{, #2#1#3}{ and~#2#1#3}
\crefrangeformat{section}{sections~#3#1#4\,--\,#5#2#6}
\crefformat{subsection}{section~#2#1#3}
\crefmultiformat{subsection}{sections~#2#1#3}
    { and~#2#1#3}{, #2#1#3}{ and~#2#1#3}
\crefrangeformat{subsection}{sections~#3#1#4\,--\,#5#2#6}
\crefformat{subsubsection}{section~#2#1#3}
\crefmultiformat{subsubsection}{sections~#2#1#3}
    { and~#2#1#3}{, #2#1#3}{ and~#2#1#3}
\crefrangeformat{subsubsection}{sections~#3#1#4\,--\,#5#2#6}
\crefformat{figure}{figure~#2#1#3}
\crefmultiformat{figure}{figures~#2#1#3}
    { and~#2#1#3}{, #2#1#3}{ and~#2#1#3}
\crefrangeformat{figure}{figures~#3#1#4\,--\,#5#2#6}
\crefformat{table}{table~#2#1#3}
\crefmultiformat{table}{tables~#2#1#3}
    { and~#2#1#3}{, #2#1#3}{ and~#2#1#3}
\crefrangeformat{table}{tables~#3#1#4\,--\,#5#2#6}
\crefformat{equation}{eq.~(#2#1#3)}
\crefmultiformat{equation}{eqs.~(#2#1#3)}{ and~(#2#1#3)}{, (#2#1#3)}{ and~(#2#1#3)}
\crefrangeformat{equation}{eqs.~(#3#1#4)--(#5#2#6)}

\renewcommand{\hookAfterAbstract}{%
\par\bigskip\bigskip
CERN-TH-2024-005
}

\begin{document}
\maketitle

\section{Introduction}

Lattice QCD is an extremely solid and successful theoretical framework and has recently become essential for precision physics calculations, as it is able to provide a number of hadronic matrix elements and Standard Model parameters with uncertainties in some cases even smaller than one percent~\cite{Aoki_2022}.
At this level of precision, small effects such as the electromagnetic~(e.m.) interactions of quarks and leptons, as well as the difference in mass between up and down quarks, which are both expected to amount to $\Ocal(1\%)$ corrections, play a crucial role.
In this work we discuss the calculation of such isospin-breaking~(IB) effects in the context of weak processes, in particular of leptonic decays of pseudoscalar mesons. These processes are mediated by charged-weak currents and hence the combination of a precise experimental determination of their decay rates and the theoretical calculation of the relevant hadronic amplitudes, including IB effects, can provide precise estimates of the corresponding Cabibbo-Kobayashi-Maskawa (CKM) matrix elements.
The unitarity contraints on the CKM matrix imposed by the Standard Model offer a unique opportunity for indirect searches of new physics. Significant deviations from unitarity could in fact hint at contributions of new particles or interactions which are not accounted in the theoretical calculations.

The evaluation of IB corrections to hadronic observables requires the inclusion of QED in numerical lattice QCD calculations, posing the theoretical challenge of how to define a photon in a finite volume with periodic boundary conditions, which are commonly employed in lattice calculations. In fact, Gauss' law forbids states with non-zero electric charge in a box with periodic boundary conditions. In order to circumvent this problem, many regularizations of the lattice QED action have been proposed, that either put contraints on the Fourier modes of the photon field, like $\QEDL$~\cite{Hayakawa:2008an} and its infrared improvements~\cite{Davoudi:2018qpl}, $\QEDTL$~\cite{Duncan:1996xy} and $\QEDSF$~\cite{GOCKELER1990527}, or provide a small mass to the photon like $\QEDm$~\cite{Endres:2015gda}, or adopt instead different boundary conditions like $\QEDC$~\cite{Lucini:2015hfa}. For a detailed review discussing all the above prescriptions, see ref.~\cite{Patella:2017fgk}. A different alternative approach has also been recently developed, in which radiative corrections are determined as a convolution of hadronic correlators with infinite-volume QED kernels~\cite{Feng:2018qpx,Christ:2023lcc}. In this case, effects due to the finite extent of the lattice are expected to come only from QCD hadronic matrix elements, and hence decay exponentially as the lattice extent~$L$ is increased. In contrast, when QED is formulated in a finite volume with a massless photon, the long-range nature of e.m.~interactions leads to power-like finite-volume effects.  In this work we will focus on lattice calculations in which QED has been defined in a finite volume and regularized using the $\QEDL$ prescription, which consists in removing the contributions of the spatial zero modes of the photon, and we will put particular emphasis on the crucial role played by such power-like finite-volume corrections.

There exist two approaches currently adopted to include QED effects in lattice QCD calculations. In the all-order or ``stochastic'' approach QED is added directly to the lattice action, and dedicated QCD+QED simulations are performed, with the result of including the IB corrections to all orders (see, e.g., refs.~\cite{Borsanyi_2015,Boyle:2017gzv,Hansen:2018zre,RCstar:2022yjz}). In the perturbative approach, presented in refs.~\cite{deDivitiis:2011eh,deDivitiis:2013xla} (and  often referred to as the ``RM123~approach''), the lattice path-integral is expanded at first order in the two small parameters $\aem$ and $(m_\mathrm{d}-m_\mathrm{u})/\Lambda_\QCD$. Being leading order corrections sufficient for many phenomenological applications, the two small parameters are factorized out and their coefficients can be determined directly from simulations of isosymmetric QCD, with no need to perform new dedicated simulations. So far, due to the computational challenge of evaluating quark disconnected diagrams, all calculations using the perturbative method have been performed in the electro-quenched approximation, in which sea quarks are treated as electrically neutral. Although deviations from this approximation are expected to be small, this consitutes a major source of systematic uncertainty and therefore work is in progress to overcome it~\cite{Harris:2023zsl}.

Lattice QCD+QED is a new frontier for precise numerical calculations and many collaborations have successfully produced results for IB corrections to different hadronic observables: the hadron spectrum~\cite{Borsanyi_2015,Fodor_2016,Giusti:2017dmp,Feng:2021zek,Frezzotti:2022dwn}, the anomalous muon~$g{-}2$~\cite{RBC:2018dos, Giusti:2019xct, Borsanyi:2020mff, Ce:2022kxy, Biloshytskyi:2022ets, Chao:2023lxw}~(see also refs.~\cite{Kuberski:2023,Aoyama:2020ynm}) and weak leptonic decays~\cite{Giusti:2017dwk,DiCarlo:2019thl,Boyle_2023}. In the latter case, additional challenges arise. In fact, when separating the contributions to the decay rate from the exchange of virtual photons and those from the emission of real ones, logarithmic infrared divergences appear which have to be properly treated.
The strategy for evaluating leading e.m.~and strong IB effects to leptonic decay rates on the lattice has been developed by the RM123-Southampton (RM123S) collaboration in ref.~\cite{Carrasco:2015xwa} and then applied by the same group to the leptonic decay of pions and kaons into muons and neutrinos (referred to as $\pmuTwo$ and $\KmuTwo$, respectively) in refs.~\cite{Giusti:2017dwk,DiCarlo:2019thl}. More recently, the RBC/UKQCD collaboration also computed the IB correction to the ratio of kaon and pion decay rates, $\Gamma(\KmuTwo)/\Gamma(\pmuTwo)$, obtaining results in agreement with RM123S, but with a larger systematic uncertainty related to finite-volume effects. The finite-volume dependence of this observable has been thoroughly studied in the last few years. The structure-independent logarithmic infrared divergence and the e.m.~finite-volume effects to decay amplitudes up to $\Ocal(1/L^2)$  have been first computed in refs.~\cite{Lubicz:2016xro,Tantalo:2016vxk}. More recently, the structure-dependent $\Ocal(1/L^2)$ corrections, as well as the point-like contribution at $\Ocal(1/L^3)$, have been evaluated in ref.~\cite{DiCarlo:2021apt}, while the functional form of the structure-dependent effects at $\Ocal(1/L^3)$ has been presented at this conference in ref.~\cite{DiCarlo:2023rlz}.

In this work, after briefly describing how to compute IB corrections to leptonic decays on the lattice and evaluate the corresponding leading finite-volume effects, we will discuss the two existing calculations of the ratio $\Gamma(\KlTwo)/\Gamma(\plTwo)$ performed by the RM123S and the RBC/UKQCD groups using the $\QEDL$ regularization. We will focus on the role of finite-volume effects in the calculations and how further studies of the correction at $\Ocal(1/L^3)$ are necessary to improve the precision on such observable.
With this aim, we dedicate the rest of the work to present a new finite-volume regularization of the QED action, which differs from $\QEDL$ in the treatment of the photon zero mode and it is expected to reduce finite-volume effects on a number of hadronic observables. 

\section{QED and isospin-breaking corrections to weak decays}
\label{sec:weak_decays}

Let us consider the decay of a charged pseudoscalar meson, $P^+$, into a muon-neutrino pair. When electromagnetism is included, photons can be exchanged between quarks and leptons or emitted by them. As mentioned above, this introduces new infrared divergences in the decay rate, which only cancel when summing virtual ($\Gamma_0$) and real ($\Gamma_1$) photon contributions~\cite{Bloch:1937pw}. 
A practical strategy for lattice calculations of the decay rate, including leading QED corrections, has been proposed by RM123S in ref.~\cite{Carrasco:2015xwa} and consists in defining the decay rate as the sum of infrared-finite contributions, namely
\begin{align}
    \Gamma(P^+\to \mu^+\nu_\mu[\gamma]) &= 
    \lim_{L\to \infty} \big[\Gamma_0(L)-\Gamma_0^\mathrm{uni}(L)\big] + \lim_{m_\gamma\to \infty} \big[\Gamma_0^\mathrm{uni}(m_\gamma)+\Gamma_1^\mathrm{uni}(m_\gamma)\big]\\
    & +\lim_{L\to \infty} \big[\Gamma_1(L)-\Gamma_1^\mathrm{uni}(L)\big]\,. \nn
\end{align}
In the first bracketed term the virtual decay rate is computed on the lattice using the finite volume with the $\QEDL$ prescription as an IR regulator, with the universal (structure-independent) logarithmic infrared divergence  removed, as well as finite volume effects up to $\Ocal(1/L)$~\cite{Lubicz:2016xro,Tantalo:2016vxk}. The finite-volume scaling of this term can be further improved by replacing $\Gamma_0^\mathrm{uni}(L)$ with $\Gamma_0^{(2)}(L)$, which includes the $\QEDL$ finite-volume corrections up to $\Ocal(1/L^2)$, computed in ref.~\cite{DiCarlo:2021apt}. The second term is the decay rate in the point-like approximation, which has been evaluated in perturbation theory using a photon mass as a regulator in ref.~\cite{Carrasco:2015xwa}. Finally, the third term corresponds to the structure-dependent part of the real decay rate. This contribution has been studied in refs.~\cite{deDivitiis:2019uzm,Kane:2019jtj,Kane:2021zee,Frezzotti:2020bfa,Giusti:2023pot,Desiderio:2020oej,Frezzotti:2023ygt} and it is relevant for decays of pions and kaons into electrons or decays of heavy mesons, while it can be neglected in the calculations of $\pmuTwo$ and $\KmuTwo$. 
This method has been applied to two different numerical calculations, first by the RM123S group to the rates of $\pmuTwo$ and $\KmuTwo$~\cite{Giusti:2017dwk,DiCarlo:2019thl}, and more recently by the RBC/UKQCD collaboration to the ratio $\Gamma(\KmuTwo)/\Gamma(\pmuTwo)$~\cite{Boyle_2023}. In the following we focus on the latter quantity and compare it between the two calculations. 

For a given choice of scheme for the iso-symmetric QCD theory ($\aem=0$, $m_\mathrm{u}=m_\mathrm{d}$), in which the decay constants $f_\pi$ and $f_K$ are defined, one can write the ratio of $\KmuTwo$ and $\pmuTwo$ decay rates as 
\begin{equation}
    \frac{\Gamma(K^+\to \mu^+\nu_\mu[\gamma])}{\Gamma(\pi^+\to \mu^+\nu_\mu[\gamma])} = \frac{|V_\mathrm{us}|^2}{|V_\mathrm{ud}|^2}\frac{m_\pi}{m_K}\frac{(m_K^2-m_\mu^2)}{(m_\pi^2-m_\mu^2)}\frac{f_K^2}{f_\pi^2}\big(1+\delta R_{K\pi}\big) + \Ocal(\epsilon^2)\,,
\end{equation}
where $\delta R_{K\pi}=\delta R_K-\delta R_\pi$ denotes the (scheme-dependent) leading IB correction and $\Ocal(\epsilon^2)$ is understood as a second-order correction in $(\aem, (m_\mathrm{d}^\mathrm{R} - m_\mathrm{u}^\mathrm{R} )/\Lambda_\QCD)$.%
\footnote{We stress that the definition of the isospin symmetric QCD theory is scheme dependent, as is the separation of strong and e.m. IB effects. Although differences between commonly adopted schemes are expected to be small, discussions are ongoing for the definition of a common reference scheme, which would facilitate the comparison of lattice results between different collaborations~\cite{Tantalo:2023onv}. As discussed in ref.~\cite{Boyle_2023}, the schemes adopted in the two calculations discussed in this work can be considered to be equivalent.}
As discussed in refs.~\cite{Giusti:2017dwk,DiCarlo:2019thl,Boyle_2023}, the IB correction $\delta R_{K\pi}$ can be obtained by computing corrections to the bare matrix elements and to the meson masses, which can be extracted from the long-distance behaviour of suitable Euclidean lattice correlators in the time-momentum representation. Note that $\delta R_{K\pi}$ depends on the  the velocities of the final state leptons in the pion and kaon decays, whose absolute value is fixed by energy-momentum conservation and injected in the lattice correlators via twisted boundary conditions. In both calculations the RM123 perturbative method has been adopted to evaluate QED and strong IB effects in the electro-quenched approximation. This consists, in practice, in computing a number of connected correlation functions, with photons exchanged in all possible ways and mass corrections inserted along all quark lines.

The lattices used in the two calculations are rather different. On the one hand, RM123S employed ensembles with twisted mass fermions at three different lattice spacings and multiple volumes. However, unphysical quark masses have been simulated, corresponding to pion masses above 230~MeV, and a chiral extrapolation has been performed to reach the physical point. One complication introduced by the use of an action that breaks chiral symmetry is a non-trivial mixing of operators under renormalization when including QED corrections. This has been studied in ref.~\cite{DiCarlo:2019thl}, but the issue does not apply to the calculation of $\delta R_{K\pi}$, as the contributions from the renormalization of the matrix elements cancel in the ratio. On the other hand, the RBC/UKQCD calculation used the domain wall fermion action with physical quark masses. However, due to the high computational cost of such simulation, this has been performed for the moment at a single value of lattice spacing and volume.

The results obtained by the two collaborations are the following~\cite{DiCarlo:2019thl,Boyle_2023}
\begin{equation}
    \delta R_{K\pi}^\mathrm{RM123S} = -0.0126\,(21)\,,
    \qquad
    \delta R_{K\pi}^\mathrm{RBC/UKQCD} = -0.0086\,(13)(39)_\mathrm{vol.}\,,
\end{equation}
where the error in the first bracket is a combination of statistical and systematic uncertainties, while for the RBC/UKQCD result we factor out the systematic error due to finite-volume effects. The two results are in agreement with each other, as well as with a previous calculation obtained in chiral perturbation theory,~$\delta R_{K\pi}^\mathrm{\chi PT}=-0.0112\,(21)$~\cite{Cirigliano:2011tm}. This is a noteworthy result, emphasizing the  capability of the lattice in reliably computing such observables with a precision systematically improvable to the percent level.

The origin of the large finite-volume systematic uncertainty in the RBC/UKQCD result is explained in ref.~\cite{Boyle_2023} and additional details are also given in ref.~\cite{DiCarlo:2023rlz}. This is due to the partial knowledge of finite-volume corrections to the decay rate at $\Ocal(1/L^3)$. While the scaling in inverse powers of $L$ is known for this quantity up to $\Ocal(1/L^2)$, including structure-dependent contributions, the $\Ocal(1/L^3)$ term is only known in the point-like approximation. It can be shown that the structure-dependent part receives contributions from branch-cuts in hadronic amplitudes, which are difficult to evaluate numerically. The appearance of these contributions is a combined effect of the breaking of spatial locality, due to the use of the $\QEDL$ action, and of rotational symmetry breaking in a finite volume, as it will be discussed in the next section. In addition, the known point-like contribution at $\Ocal(1/L^3)$ is found to be sizeable (and with opposite sign) compared to the lower order corrections, thus giving rise to a final large value of the systematic uncertainty. 
It is worth mentioning that the systematic error assigned by RBC/UKQCD is likely over-conservative, while the RM123S one might be slightly underestimated. With the improved understanding of finite-volume corrections to $\delta R_{K\pi}$ since the publication of the RM123S result, an updated determination is certainly within reach for the RM123S group.
Reducing the finite-volume systematic error requires a better knowledge of the finite-volume scaling of $\delta R_{K\pi}$. This can be studied with a combination of numerical and analytical calculations, for example by repeating the calculation on a number of different volumes and extrapolating the data to the infinite-volume limit (undoubtedly the cleanest but most expensive approach), while trying to estimate the missing terms in the $1/L$ expansion. In this work we propose an additional approach to improve the finite-volume scaling of leptonic decay rates and other hadronic observables, which consists in a modification of the finite-volume QED action. This new prescriptions, which we call $\QEDr$, leads to vanishing contributions at $\Ocal(1/L^3)$, promising a substantial improvement in the precision of $\delta R_{K\pi}$. 

\section{\texorpdfstring{$\QEDr$}{QEDr}: an infrared improved QED regularization}
\label{sec:qedr}

\begin{figure}
    \centering
    \includegraphics[width=0.9\textwidth]{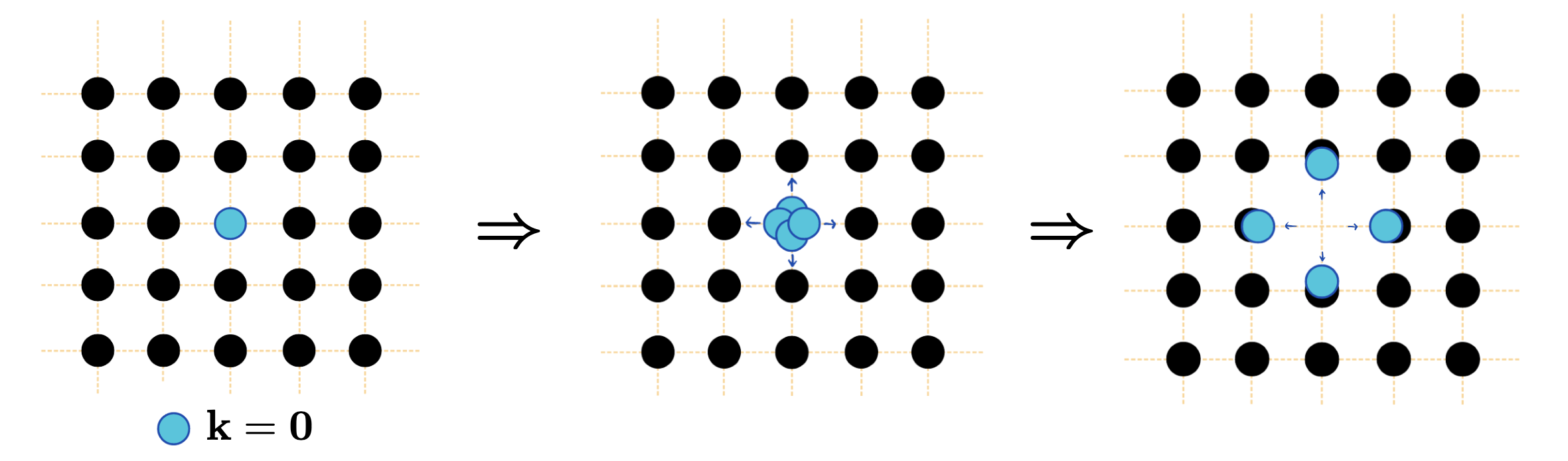}
    \caption{Visualisation of the redistribution of the spatial zero mode $\kvec=\zero$ over the nearest neighbouring modes on the shell with radius $R=1$ with equal weights.}
    \label{fig:qedr}
\end{figure}

The removal of the spatial zero modes from the photon propagator in the $\QEDL$ regularization corresponds to placing a uniform charge density in the volume, such that the finite-volume Gauss' law is respected even with a non-zero electric charge in the volume. 
As a result, hadronic observables are affected by finite-volume effects at order $1/L^3$ which, as mentioned in the previous section, are related to branch-cuts in hadronic amplitudes and therefore difficult to estimate.
We propose here a variation of the $\QEDL$ regularization that restores the zero-mode contribution in the infinite-volume limit and does not generate $\Ocal(1/L^3)$ corrections. 
This proposal can be seen as a particular case of the infrared improvement of the $\QEDL$ action introduced in ref.~\cite{Davoudi:2018qpl}, in which a finite number of photon momentum modes are reweighted to obtain a better finite-volume scaling of a given observable, without altering its infinite-volume limit. While the coefficients introduced in ref.~\cite{Davoudi:2018qpl} were improving finite-volume effects only in a process-dependent fashion, the new choice we present here allows for a universal removal of momentum-independent $1/L^3$ effects.
Our proposal, which we refer to as $\QEDr$, consists in redistributing the zero mode, $\kvec=\zero$, over the set of neighbouring modes, $\kvec\in\Scal_R$, lying on a sphere of radius~$\tfrac{2\pi}{L}R$, namely 
\begin{align}
    \Scal_R &= \big\{ \,\kvec\in\tfrac{2\pi}{L}\mathbb{Z}^3 \ \ \big| \ \ |\kvec| = \tfrac{2\pi}{L} R \big\}\,.
\end{align}
In~\cref{fig:qedr} we show a schematic visualisation of such redistribution of the zero mode onto the nearest neighour modes on the shell with radius $R=1$.

\begin{figure}
    \centering
    \includegraphics[width=\textwidth]{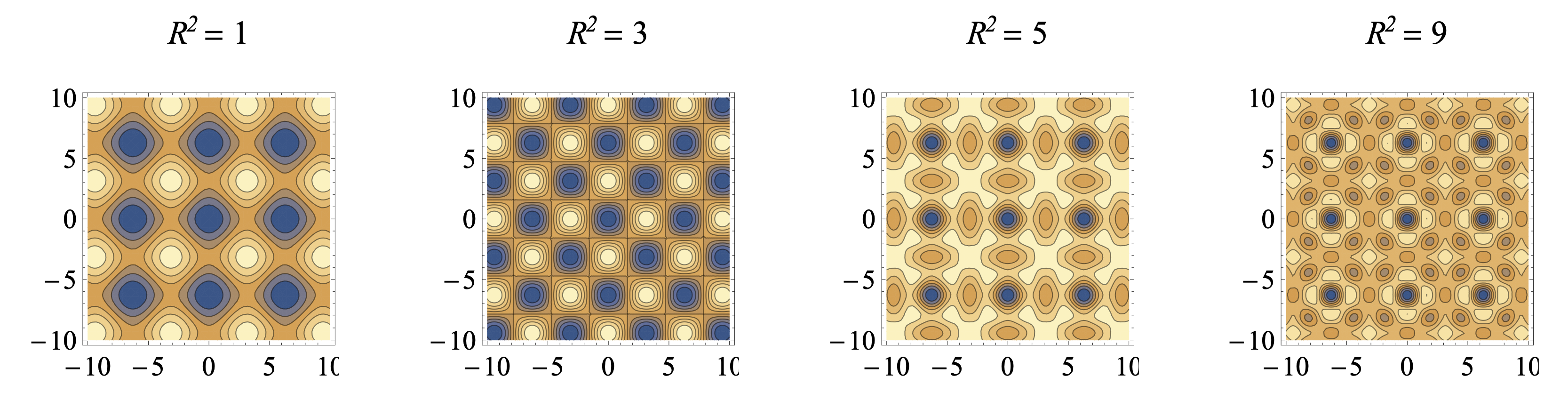}
    \caption{Contour plot of the charge density for different choices of the shell radius $R$.}
    \label{fig:qedr_charge_densities}
\end{figure}

The infrared-improved Euclidean $\QEDr$~propagator is implemented, in practice, by adding to the $\QEDL$ propagator the contributions of the modes $\kvec\in \Scal_R$. Denoting $\kvec = \tfrac{2\pi}{L}\nvec$, the $\QEDr$~propagator in Feynman gauge reads
\begin{equation}
    D^{\mu\nu}(k_0,\kvec) = \delta^{\mu\nu}\ \frac{1-\delta_{\kvec,\zero}}{k_0^2+\kvec^2}\ \left[1+h(\kvec,R)\right]
    \label{eq:prop_qedr}
\end{equation}
where the first term in the sum corresponds to the $\QEDL$ propagator and
\begin{equation}
    \label{eq:weights}
    h(\kvec,R) = w(\kvec,R)\,\delta_{\nvec^2,R^2}\,.
\end{equation}
The function $w(\kvec,R)$ denotes the weight assigned to the mode $\kvec$ on the shell $\Scal_R$ and is defined such that the sum of all weights equals one,
\begin{equation}
    \sum_{\kvec\in\Scal_R} w(\kvec,R)=1\,,
\end{equation}
which implies that
\begin{equation}
    \sum_\kvec h(\kvec,R)=1\,.
    \label{eq:sum_h}
\end{equation}
Later in this section, we will demonstrate that the condition  in~\cref{eq:sum_h} is crucial for the cancellation of the finite-volume corrections at $\Ocal(1/L^3)$.
In the case of an isotropic system with zero net velocity the natural choice is to assign equal weights to the modes $\kvec\in\Scal_R$ with values $w(\kvec,R)=1/r_3(R^2)$, where the function $r_3(R^2)=\sum_{\kvec}\delta_{\kvec^2,R^2}$ counts the number of representation of $R^2$ as the sum of 3 squares.
Following the definitions above, it is clear that in the $\QEDr$ regularization the electric charge density in the volume is not uniform as in $\QEDL$, but it is a function of the spatial coordinates with a periodicity that depends on the radius~$R$, as shown in~\cref{fig:qedr_charge_densities}.

In principle, the definition of the propagator in~\cref{eq:prop_qedr} can be extended to the case of the zero mode redistributed over multiple shells with radii $\boldsymbol{R}=\{R_1,R_2,\dots\}$ as follows
\begin{equation}
    D^{\mu\nu}(k_0,\kvec) = \delta^{\mu\nu} \ \frac{1-\delta_{\kvec,\zero}}{k_0^2+\kvec^2} \ \left[1+h(\kvec,\boldsymbol{R})\right]\,,
\end{equation}
where this time
\begin{equation}
    h(\kvec,\boldsymbol{R}) = \sum_{\alpha}\omega(\alpha) \, h(\kvec,\alpha)\,,
\end{equation}
with $\omega(\alpha)$ denoting the weight of the shell $|\nvec|=\alpha$ and defined such that $\sum_{\alpha}\omega(\alpha)=1$. 

In the following, we will focus on the simplest implementation of $\QEDr$, corresponding to a redistribution of the zero mode over a single shell~$\Scal_R$ with radius~$R$ with all the modes equally weighted, namely with $h(\kvec,R)=\delta_{\nvec^2,R^2}/r_3(R^2)$. In this case, the propagator reads
\begin{equation}
    D^{\mu\nu}(k_0,\kvec) = \delta^{\mu\nu}\ \frac{1}{k_0^2+\kvec^2}\ \left[(1-\delta_{\kvec,\zero})+ h(\kvec,R)\right] \equiv D_\mathrm{L}^{\mu\nu}(k_0,\kvec) + D_\mathrm{r}^{\mu\nu}(k_0,\kvec)\,,
    \label{eq:prop_qedr_simple}
\end{equation}
where in the second equation we have identified the $\QEDL$ propagator and the $\QEDr$ correction.

We now turn to discuss the impact of the $\QEDr$ infrared improvement on the finite-volume scaling of hadronic observables. 
We first study observables with no dependence on an external spatial momentum. This is the case, for instance, of the e.m.~corrections to hadron masses or to the HVP contribution to the anomalous muon $g{-}2$. Then, we will discuss the more complicated scenario of a system that depends on some external non-zero spatial momentum, like the case of IB corrections to leptonic decays, where the decay amplitude depends on the velocity of the final-state lepton.

\subsection{Momentum-independent observables}

Let us consider the case of e.m.~finite-volume corrections to the mass of a charged hadron. These have been studied in details in the~$\QEDL$ regularization in refs.~\cite{Borsanyi_2015,Davoudi:2014qua,DiCarlo:2021apt}, as well as in~$\QEDC$~\cite{Lucini:2015hfa}. We find it convenient to adopt here the notation of ref.~\cite{Lucini:2015hfa}, to which we refer for further discussions on the analytical properties of the functions appearing below.

In $\QEDr$, finite-volume corrections to a charged hadron mass can be written as
\begin{equation}
    \Delta m^2(L) = \Delta m_\mathrm{L}^2(L) + \Delta m_\mathrm{r}^2(L)\,,
\end{equation}
where, similarly to~\cref{eq:prop_qedr_simple}, the first term on the right hand side corresponds to the $\QEDL$ correction, while the second term is the additional contribution from the redistributed zero mode.

The $\QEDL$ e.m. finite-volume effects can then be obtained as follows
\begin{equation}
    \Delta m^2_\mathrm{L}(L) = e^2 \, \Delta_\kvec' \int\frac{\dd k_0}{2\pi} \frac{{{M_{\mu}}^\mu}(\ii k_0,\kvec)}{k_0^2+\kvec^2} = \frac{e^2}{2} \,  \Delta_\kvec' \frac{{{M_{\mu}}^\mu}(-|\kvec|,\kvec)}{|\kvec|}\,,
    \label{eq:deltamL}
\end{equation}
where $\Delta_\kvec' = \Big[\tfrac{1}{L^3}\sum_{\kvec\neq \boldsymbol{0}}-\int\tfrac{\dd^3 \kvec}{(2\pi)^3}\Big]$ and
\begin{equation}
    {M_\mu}^\mu(\ii k_0,\kvec) = \frac{Z_\mathrm{1P}(\kvec^2)}{\sqrt{m^2+\kvec^2}-m - \ii k_0} + Z_\mathrm{MP}(\ii k_0,\kvec^2)
\end{equation}
is the retarded Compton tensor, split into single-particle (1P) and multi-particle (MP) contributions. Once evaluated at $(-|\kvec|,\kvec)$, it can be written as
\begin{equation}
    {M_\mu}^\mu(-|\kvec|,\kvec) = \frac{Z_\mathrm{1P}(0)}{|\kvec|} + \mathcal{M}(|\kvec|)\,,
\end{equation}
with $\mathcal{M}(|\kvec|)$ being a regular function of $|\kvec|$.
Substituting  $|\kvec|=\tfrac{2\pi}{L}|\nvec|$ in~\cref{eq:deltamL} and expanding for large values of $L$, one obtains
\begin{equation}
    \Delta m^2_\mathrm{L}(L) = \frac{e^2}{2}\,\bigg[ c_1\frac{ Z_\mathrm{1P}(0)}{4\pi L} + c_2 \, \frac{\mathcal{M}(0)}{2\pi^2 L^2} +c_0\, \frac{\mathcal{M}'(0)}{L^3} -\frac{1}{2\pi^2} \sum_{\ell=0}^\infty \frac{(-1)^{\ell}\, c_{4+2\ell}}{L^{4+2\ell}}   \, \mathcal{M}^{(2+2\ell)}(0)\bigg]\ ,
    \label{eq:mass_FVE_QEDL}
\end{equation}
where the finite-volume coefficients 
\begin{equation}
    c_j = \Big[\sum_{\nvec\neq\zero} - \int \dd^3\nvec \Big] \frac{1}{|\nvec|^j} 
\end{equation}
have been introduced in ref.~\cite{Davoudi:2014qua} and discussed in refs.~\cite{Davoudi:2018qpl,DiCarlo:2021apt}.
We stress that the finite-volume correction at $\Ocal(1/L^3)$ does not vanish in $\QEDL$ since $c_0 = -1$. While the terms $Z_\mathrm{1P}(0)$ and $\mathcal{M}(0)$ are universal and only depend on charge and mass of the hadron, the contribution at $\Ocal(1/L^3)$ and beyond depend also on the internal structure of the particle. As discussed in ref.~\cite{DiCarlo:2021apt}, evaluating the contribution denoted here as $\Mcal'(0)$ requires computing an integral over the branch cut of the forward Compton amplitude, which makes the estimation of the $\Ocal(1/L^3)$ correction rather complicated. However, we can prove now that such effect is cancelled in $\QEDr$. In fact, the additional contribution to the e.m.~finite-volume effects to the hadron mass is obtained by following similar steps and evaluating
\begin{align}
    \Delta m^2_\mathrm{r}(L)= e^2 \, \sum_{\kvec\neq \boldsymbol{0}} \, h(\kvec,R)\int\frac{\dd k_0}{2\pi} \frac{{{M_{\mu}}^\mu}(\ii k_0,\kvec)}{k_0^2+\kvec^2} = \left.\frac{e^2}{2} \,  \frac{{{M_{\mu}}^\mu}(-|\pvec|,\pvec)}{|\pvec|}\right|_{|\pvec|=2\pi R/L}  \,.
    \label{eq:Cottingham_eucl_new}
\end{align}
Substituting now  $|\pvec|=\tfrac{2\pi}{L}R$ and expanding again for large values of $L$ yields
\begin{align}
    \Delta m^2_\mathrm{r}(L) &= \frac{e^2}{2}\,\bigg[ \frac{R^{-2}}{\pi}\frac{ Z_\mathrm{1P}(0)}{4\pi L} + (\pi R^{-1} ) \, \frac{\mathcal{M}(0)}{2\pi^2 L^2} + \frac{\mathcal{M}'(0)}{L^3} \, + \\
    &\qquad \qquad \quad + \sum_{\ell=0}^\infty \frac{(2\pi R)^{2\ell+1}}{(2\ell+2)!\, L^{4+2\ell}} \mathcal{M}^{(2\ell+2)}(0) + \sum_{\ell=0}^\infty \frac{(2\pi R)^{2\ell+2}}{(2\ell+3)!\, L^{5+2\ell}} \mathcal{M}^{(2\ell+3)}(0) \bigg]\,.\nn
\end{align}
Combining this result with the $\QEDL$ correction in~\cref{eq:mass_FVE_QEDL} we obtain the full e.m.~finite-volume correction to the mass
\begin{align}
    \Delta m^2(L) &= \frac{e^2}{2}\,\bigg[ \bar{c}_1(R)\frac{ Z_\mathrm{1P}(0)}{4\pi L} + \bar{c}_2(R) \, \frac{\mathcal{M}(0)}{2\pi^2 L^2} +\bar{c}_0\, \frac{\mathcal{M}'(0)}{L^3} +\\
    &\qquad\qquad\quad  -\frac{1}{2\pi^2} \sum_{\ell=0}^\infty \frac{(-1)^{\ell}\, \bar{c}_{4+2\ell}(R)}{L^{4+2\ell}} \mathcal{M}^{(2\ell+2)}(0) + \sum_{\ell=0}^\infty \frac{\bar{c}_{5+2\ell}(R)}{L^{5+2\ell}} \, \mathcal{M}^{(3+2\ell)}(0)\bigg]\,, \nn
\end{align}
where
\begin{align}
    &\boxed{\bar{c}_0 = c_0+1 = 0}\,,\quad
    \bar{c}_1(R) = c_1 + \frac{R^{-2}}{\pi}\,,\quad
    \bar{c}_2(R) = c_2 + \pi R^{-1}\,, \nn \\
    &
    \bar{c}_{4+2\ell}(R) = {c}_{4+2\ell} + \frac{(-1)^{1-\ell}\pi (2\pi)^{2\ell+2} }{(2\ell+2)!}R^{1+2\ell}\,,\\
    &
    \bar{c}_{5+2\ell}(R) = \frac{(2\pi)^{2\ell+2}}{(2\ell+3)!}R^{2\ell+2}\,.
\end{align}
Some comments on these results are in order. First, we note that the coefficient associated with the zero mode vanishes, namely $\bar{c}_0=0$. As anticipated below~\cref{eq:sum_h}, this is a direct consequence of the constraint on the weights $\sum_\kvec h(\kvec,R)=1$, which characterizes $\QEDr$ among all possible infrared improvements of the QED action~\cite{Davoudi:2018qpl}. We can interpret this result as follows: as the volume increases, the Fourier space becomes denser and, in the infinite-volume limit, the redistributed modes ``reproduce'' the zero-mode contribution, which in $\QEDL$ would be simply removed.
Then, we note that the other $\QEDL$ finite-volume coefficients are shifted by an amount that depends on the radius $R$. For these coefficients, choosing the shell $R=1$ ($|\pvec|={2\pi}/{L}$ and $w(\kvec,1)=1/6$) seems to be the optimal choice. This implementation of $\QEDr$ is the one we consider \emph{standard} and that we adopt in the rest of the work.
Finally, we observe that new (but more suppressed) finite-volume effects arise at $\Ocal(1/L^5)$ and higher odd inverse powers of~$L$. This phenomenon is associated with the redistribution of the zero mode and to the fact that spatial locality is still broken in $\QEDr$, as it is in $\QEDL$. In fact, in a local finite-volume formulation of QED, like $\QEDC$, such contributions would not arise~\cite{Lucini:2015hfa}. Nonetheless, the emergence of these new, unknown, higher-order finite-volume effects does not raise any practical concerns or limitations. This is because, even at $\Ocal(1/L^4)$, our knowledge of structure-dependent finite-volume effects is incomplete. The notable advantage of $\QEDr$ over $\QEDL$ is that systematic uncertainties to finite-volume effects on hadron masses are pushed to a higher order, specifically to $\Ocal(1/L^4)$. And for typical lattice extents of $m_\pi L\gtrsim 4$, the size of such residual power-like finite-volume effects might be comparable with that of exponentially suppressed finite-volume effects, which are commonly neglected.

In this section, we have focused on hadron masses to illustrate the impact of $\QEDr$ on observables that do not depend on an external momentum. However, similar conclusions can be drawn for other quantities, such as the e.m.~corrections to the HVP contribution to the anomalous muon $g{-}2$. 
The e.m. finite-volume corrections to the two-pion contribution to the HVP, computed in $\QEDL$ within the point-like approximation and in the photon rest frame, are detailed in ref.~\cite{Bijnens:2019ejw} and amount to
\begin{equation}
    \Delta \hat{\Pi}(q_0^2) = \frac{c_0}{(m_\pi L)^3} \, \Omega\big({q_0^2}/{m_\pi^2}\big) + \Ocal\Big(\frac{1}{L^4}\Big)\,,
\end{equation}
where $\Omega(z)$ is a dimensionless function. From this result we deduce that $\QEDr$~finite-volume corrections to the $\pi\pi$ contribution to the HVP only start at $\Ocal(1/L^4)$. 
As the inclusion of IB corrections becomes essential for achieving high precision in lattice calculations of the muon $g{-}2$ (see ref.~\cite{Kuberski:2023} for a recent review), the use of $\QEDr$ can be beneficial in mitigating systematic uncertainties related to e.m. finite-volume effects.

\subsection{Momentum-dependent observables}

Let us study now the case of an observable that depends on an external spatial momentum. We consider here the leptonic decay rate of a meson, which depends on the velocity $\vvec_\ell = \pvec_\ell/\omega_\ell$ of the final charged lepton. As discussed in~\cref{sec:weak_decays}, the calculation of the leading IB effects to this quantity suffers from a large systematic uncertainty due to the only partial knowledge of its finite-volume corrections of $\Ocal(1/L^3)$.
In this case, the realization of the $\QEDr$ improvement is much less straightforward and not automatic. This is due to the appearance in the finite-volume expansion of the decay rate of finite-volume coefficients like
\begin{equation}
    c_j(\vvec) = \bigg[\sum_{\nvec\neq\zero}-\int \dd^3 \nvec\bigg] \, \frac{1}{|\nvec|^j (1-\vvec\cdot \hat{\nvec})}\,.
\end{equation}
As shown in ref.~\cite{DiCarlo:2023rlz}, the $\QEDL$~finite-volume correction of $\Ocal(1/L^3)$ to leptonic decay rates contains both terms proportional to $c_0$ and terms proportional to $c_0(\vvec)$. While the former are cancelled in $\QEDr$ as an effect of the condition $\sum_\kvec h(\kvec,R)=1$, the latter are not, since the coefficient 
\begin{equation}
    \bar{c}_0(\vvec) = {c}_0(\vvec) + \sum_\kvec \frac{h(\kvec,R)}{1-\vvec\cdot\hat{\nvec}}
    \label{eq:c0vl_bar}
\end{equation}
is not necessarily zero. 
As studied in ref.~\cite{Davoudi:2018qpl}, the dependence of the $\QEDL$ coefficients $c_j(\vvec)$ on the direction $\hat{\vvec}$ is a direct consequence of rotational symmetry breaking on a lattice. In fact, these coefficients can be rewritten as
\begin{equation}
    c_j(\vvec) = \frac{\mathrm{arctanh}(|\vvec|)}{|\vvec|} \, c_j \, +\, f_j(\vvec)\,,
\end{equation}
with the functions $f_j(\vvec)$ encoding direction-dependent corrections, which get stronger as $|\vvec|\to 1$ and vanish once averaged over the solid angle of $\vvec$, namely $\tfrac{1}{4\pi}\int \dd \Omega_{\vvec} \, f_j(\vvec) = 0$\,. This implies that $c_0(\vvec)$ is proportional to $c_0$, up to rotational breaking effects.
Since the additional $\QEDr$ correction in~\cref{eq:c0vl_bar}, once averaged over the directions of $\vvec$, yields 
\begin{equation}
    \frac{1}{4\pi} \sum_\kvec  \int \dd \Omega_{\vvec} \, \frac{h(\kvec,R)}{1-\vvec\cdot\hat{\nvec}} = \frac{\mathrm{arctanh}(|\vvec|)}{|\vvec|} \sum_\kvec h(\kvec,R) = \frac{\mathrm{arctanh}(|\vvec|)}{|\vvec|}\,,
\end{equation}
then, considering that $c_0=-1$, the $\QEDr$ coefficient $\bar{c}_0(\vvec)$ is zero up to rotational breaking effects.
\begin{figure}[bt]
    \centering
    \includegraphics[width=.325\textwidth]{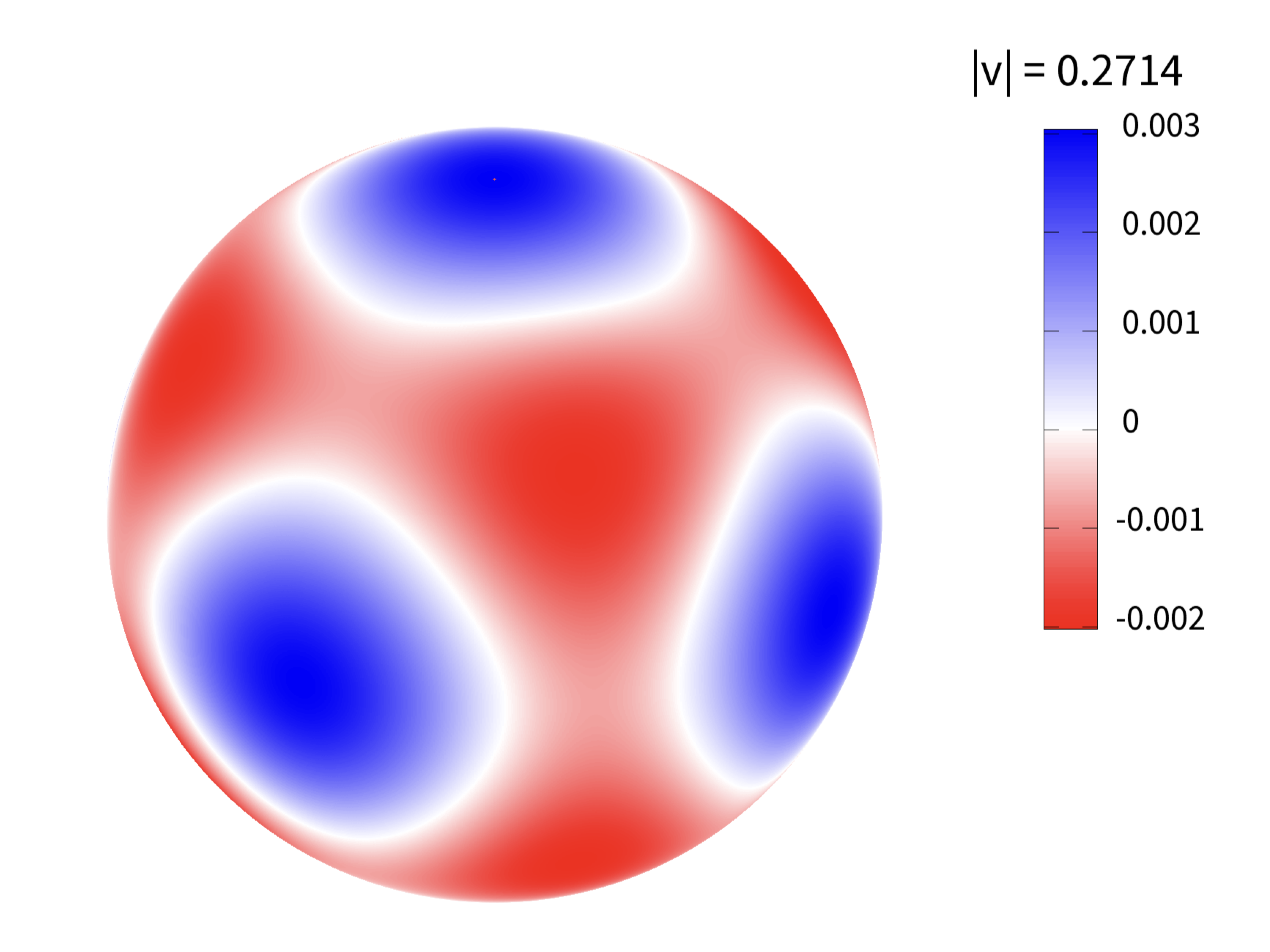}
    \includegraphics[width=.325\textwidth]{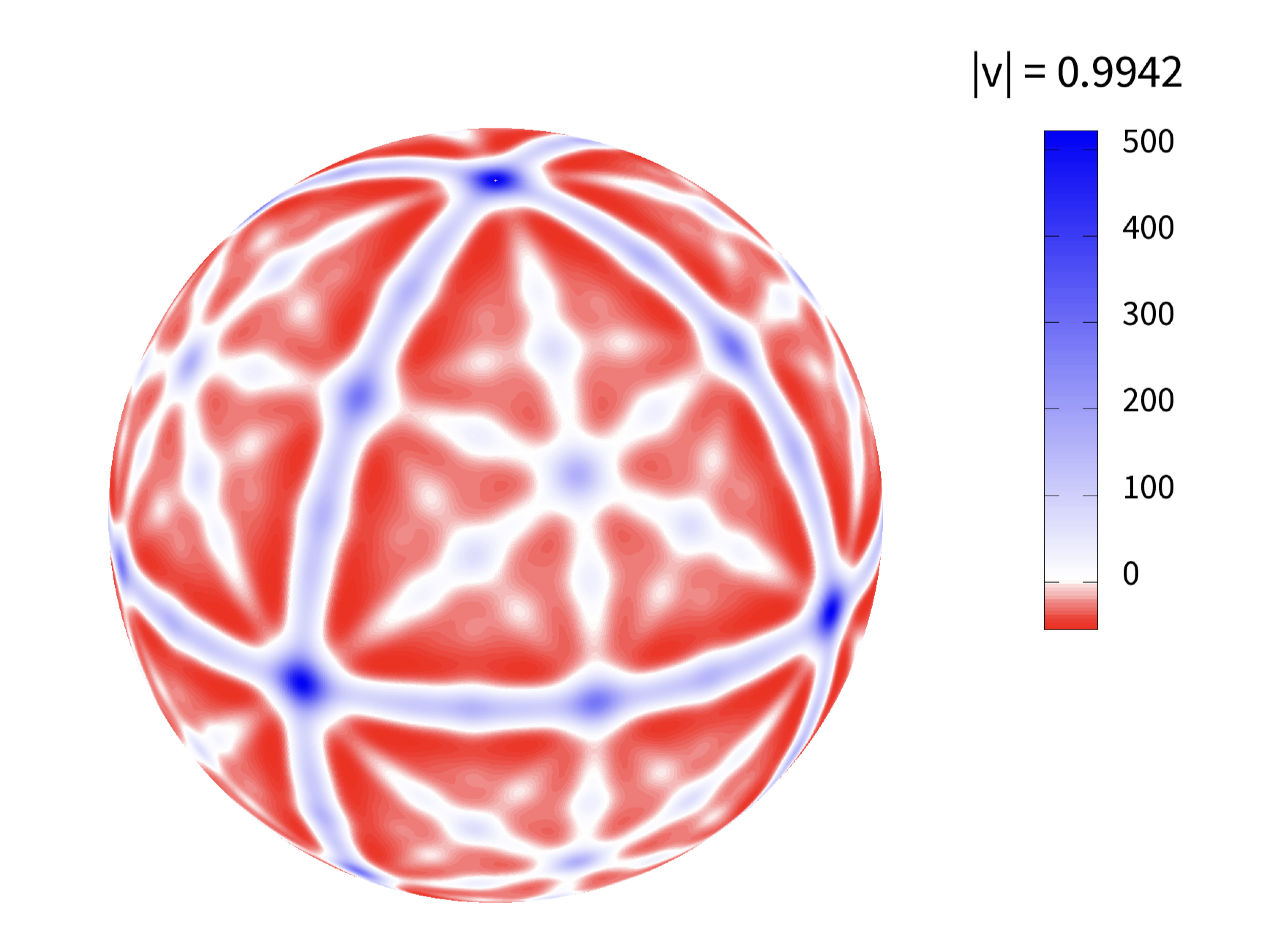}
    \includegraphics[width=.325\textwidth]{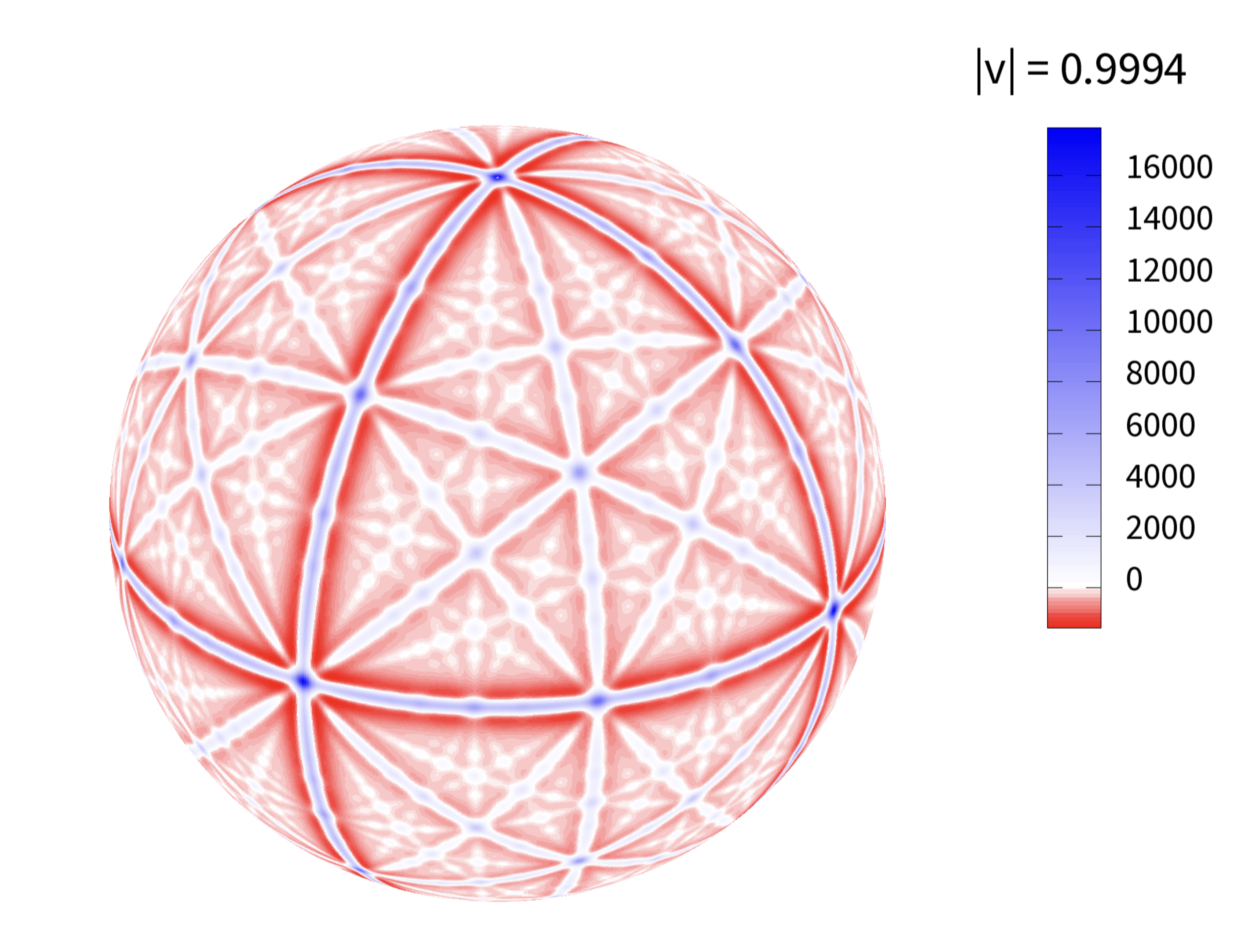}
    \caption{Angular dependence of the velocity-dependent coefficient $\bar{c}_0(\vvec)$ for three different values of $|\vvec|$. Positive and negative values of the coefficient are coloured in blue and red, respectively, while white regions correspond to the directions for which $\bar{c}_0(\vvec)=0$. Figures are retrieved from ref.~\cite{AP_Lat2023}. }
    \label{fig:spheres}
\end{figure} 
The angular dependence of $\bar{c}_0(\vvec)$ is shown in~\cref{fig:spheres} for different values of $|\vvec|$ and for the standard implementation of $\QEDr$ with $R=1$.\footnote{The pictures reported in~\cref{fig:spheres} have been taken from A.~Portelli's talk at this conference~\cite{AP_Lat2023} and have been generated using a recently developed software that allows a fast evaluation of velocity-dependent finite-volume coefficients~\cite{QedFvCoef}.} As we see, directions always exist for which $\bar{c}_0(\vvec)=0$, corresponding to the white regions in the figure. Moreover, we note that as the velocity $|\vvec|$ increases, a non-trivial fractal pattern arises, which is likely related to number-theoretical properties of the components of the vector $\hat{\vvec}$. While $|\min\bar{c}_0(\vvec)|$ remains small,  $\max\bar{c}_0(\vvec)$ diverges as $|\vvec|\to 1$, but at the same time the corresponding positive (blue) regions in~\cref{fig:spheres} get more localized and compensated by larger (red) regions where $\bar{c}_0(\vvec)$ is negative. This visually verifies that the angular average of $\bar{c}_0(\vvec)$ vanishes.
We can then summarize two fundamental properties of $\QEDr$, which hold for any value of $|\vvec|$:
\begin{enumerate}[noitemsep,nolistsep]
    \item the average of $\bar{c}_0(\vvec)$ over the solid angle of the velocity is zero;
    \item there always exists a direction $\hat{\vvec}^\star$ such that $\bar{c}_0(\vvec^\star)=0$.
\end{enumerate}
We propose here two possible ways, each related to one of the properties above, to implement the $\QEDr$ improvement and set the coefficient~$\bar{c}_0(\vvec)$ to zero in a numerical lattice calculation of momentum-dependent observables. 

One simple approach, which is easily implementable, consists in selecting the velocity with a direction $\hat{\vvec}^\star$ such that $\bar{c}_0(\vvec^\star)=0$. This process is facilitated by utilizing the C++ code $\texttt{QedFvCoef}$~\cite{QedFvCoef}, which enables a fast evaluation of velocity-dependent finite-volume coefficients, using an auto-tuned algorithm based on that proposed in ref.~\cite{Davoudi:2018qpl}. Additionally, it offers a Python binding with a suite of useful tools, including a notebook that conducts an angle scan and identifies directions for which $\bar{c}_0(\vvec^\star)=0$. 

\begin{figure}[t]
    \centering
    \includegraphics[width=.55\textwidth]{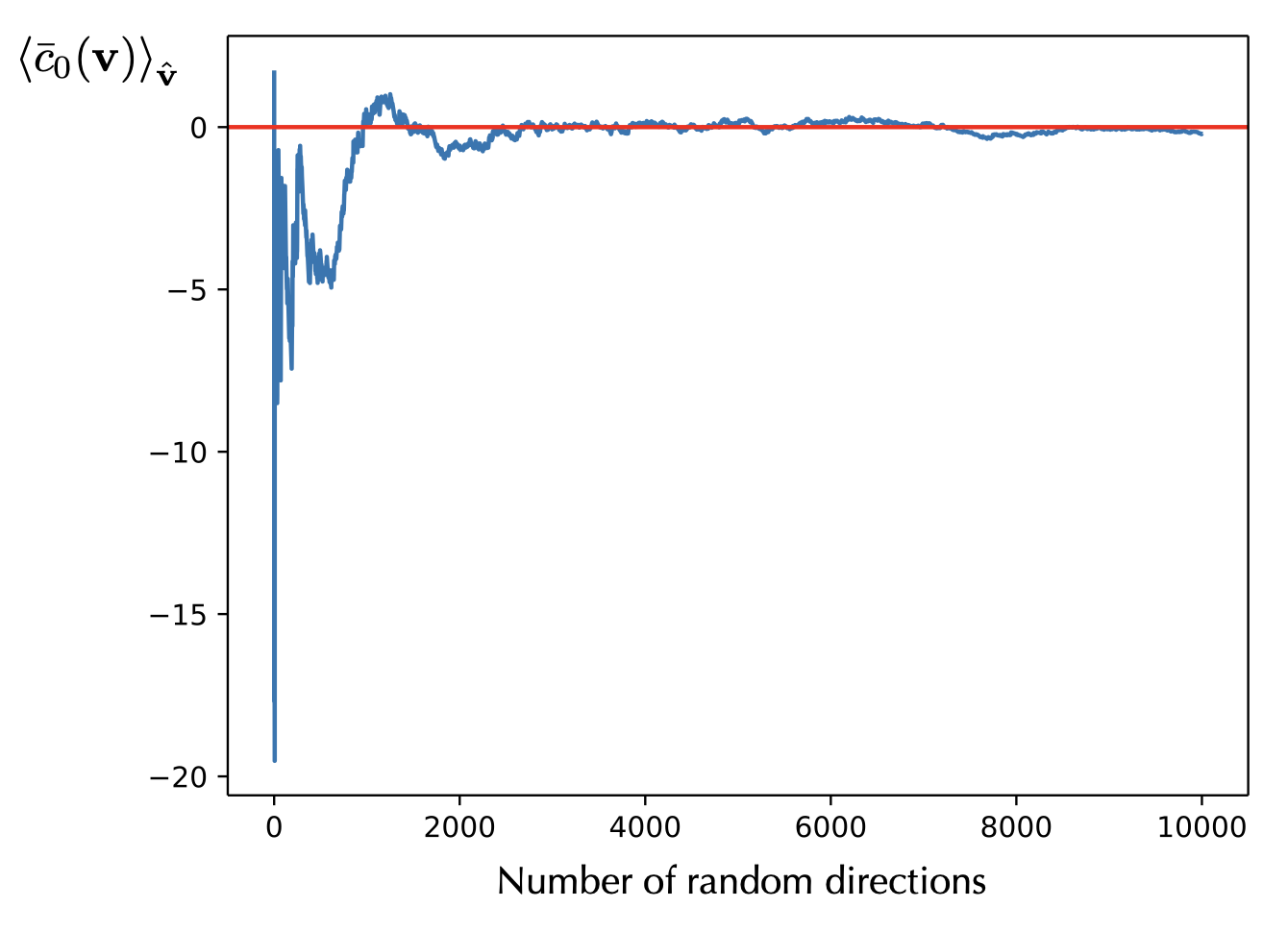}
    \caption{ Stochastic direction average of the $\QEDr$ finite-volume coefficient $\bar{c}_0(\vvec)$ at $|\vvec|=0.994$. The figure is retrieved from ref.~\cite{AP_Lat2023}. }
    \label{fig:sda}
\end{figure} 

Another possibility uses the fact that averaging over the solid angle of the velocity sets $\bar{c}_0(\vvec^\star)$ to zero. In a lattice calculation, we can achieve this result by implementing a stochastic average of the velocity directions, which are drawn randomly for each measurement. In this way, in the limit of large statistics the stochastic average 
\begin{equation}
    \langle \bar{c}_0(\vvec) \rangle_{\hat{\vvec}} = \frac{1}{N_\mathrm{meas}} \sum_{n=1}^{N_\mathrm{meas}} \bar{c}_0(\vvec_n)
\end{equation}
converges to zero. This behaviour is shown in~\cref{fig:sda}, where the stochastic average is computed for $|\vvec|=0.994$ and for an increasing number of random directions. We see that few thousands angles are needed to reproduce the desired result, the number of samples also depending on the size of the absolute value $|\vvec|$. Further details on the numerical implementation of this procedure, as well as proposals to improve the direction sampling will be discussed in a separate publication. While this procedure comes with a higher computational cost, its notable advantage lies in eliminating rotational-breaking corrections not only to the coefficient $\bar{c}_0(\vvec)$ but also to all other coefficients~$\bar{c}_j(\vvec)$.

There is, in principle, also a third way to achieve the $\QEDr$ improvement, that we briefly mention here before closing the section. While the previous two strategies use properties of $\QEDr$ defined in its standard implementation, with weights assigned in an isotropic way to the modes on the first shell of radius $R=1$, one could define the $\QEDr$ action with anisotropic weights that depend on the velocity $\vvec$ of the system studied. If the velocity is aligned along a lattice vector, namely $\vvec = \alpha \xvec$, with $\xvec=(x_1,x_2,x_3)\in \mathbb{Z}^3$ and $\mathrm{GCD}(x_1,x_2,x_3)=1$, we identify the modes on the shell of radius $R=|\xvec|$ and assign equal weight $w_1$ to the two modes along the direction of $\vvec$ and a different weight $w_2$ to all the others. We can then tune $w_1$ and $w_2$ in such a way that the following two equations are satisfied
\begin{equation}
    \begin{cases}
        \ \ \bar{c}_0(\vvec) = c_0(\vvec) + w_1 \, \Big(\frac{1}{1-|\vvec|}+\frac{1}{1+|\vvec|}\Big) + w_2 \, \sum_\kvec^\prime \frac{1}{1-\vvec\cdot\hat{\nvec}} \equiv 0 \\
        \ \ \sum_\kvec h(\kvec,R) = 2 w_1 + (r_3(R^2)-2)\, w_2 \equiv 1\,.
    \end{cases}
\end{equation}
The sum $\sum_\kvec^\prime$ in the first equation is over the modes not parallel to $\vvec$ and the second equation guarantees that $c_0=0$. Although potentially effective, the disadvantage of this implementation is that one needs to tune the lattice QED action for any specific velocity $\vvec$. Since we are interested in computing in the same simulation leptonic decay rates of different hadrons and hence with different final lepton velocities, we do not employ this implementation in our numerical studies.

\subsection{Ongoing numerical investigations}

In order to study the properties of $\QEDr$ regularization outlined in the previous sections, numerical investigations are ongoing.
On the one hand, a study at unphysical pion masses is being performed to compare the scaling of hadron masses and leptonic decay amplitudes with the volume in $\QEDL$ and $\QEDr$. Dedicated ensembles with M\"obius domain wall fermions have been generated at pion masses of around $340~\mathrm{MeV}$ and a total of four ensembles are available with lattice sizes $L/a = \{16,20,24,32\}$. Although the size of finite-volume effects depends on the pion mass, a comparison of $\QEDL$ and $\QEDr$ effects is possible also away from the physical point. Different velocities have also been simulated to test numerically the improvement in choosing a special direction such that $\bar{c}_0(\vvec^\star)=0$, to be compared with a ``naive'' one. 
On the other hand, parallel tests of the stochastic direction average are under study on a physical point domain wall fermions ensemble at a lattice spacing different from the one used in ref.~\cite{Boyle_2023}, in the context of a large scale calculation of IB corrections to decay rates of $\pi$, $K$, $D$ and $D_s$ mesons into muons and neutrinos. This calculation will also allow, in the future, to reduce systematic uncertainties on $\delta R_{K\pi}$ associated with discretization effects.
Further details on these investigations and on the ensembles used will be given in separate publications.

\section{Conclusions}

In this work we have discussed the current status of lattice calculations of IB corrections to weak decays, focusing on the two existing calculations of leptonic decay rates of pions and kaons into muons. Progress is also being made on other weak processes, and a summary of recent works in the context of kaon decays can be found in ref.~\cite{Anzivino:2023bhp}. Both calculations discussed in this work have been performed using the RM123S strategy, where the infrared divergences in the virtual and real decay rates are regularized by the introduction of an infrared cutoff. The virtual corrections to the decay rate require a non perturbative evaluation on the lattice and therefore the finite volume with the $\QEDL$ prescription for the photon action is used as an infrared regulator. The removal of the spatial zero modes of the photon in the $\QEDL$ action generates finite-volume corrections at $\Ocal(1/L^3)$, which would be otherwise absent in a local theory like $\QEDC$. While corrections to lower orders are known, the $\Ocal(1/L^3)$ contribution is only understood in the point-like approximation, ignoring the internal structure of the decaying meson. Such partial knowledge of finite-volume corrections can generate large systematic uncertainties, as observed in ref.~\cite{Boyle_2023}. In this work, we have introduced a novel approach to improve the finite-volume scaling of IB corrections in leptonic decay rates and other hadronic observables. This approach consists in a different treatment of the photon modes in the finite-volume QED action, that we call $\QEDr$. Instead of removing the spatial zero modes of the photon, $\kvec=\zero$, these are redistributed over neighbouring modes on a shell of given radius $|\kvec|=\tfrac{2\pi}{L}R$. This strategy eliminates the $\Ocal(1/L^3)$ correction, shifting the systematic uncertainty to a higher order. While this improvement occurs automatically for hadronic observables independent of external spatial momenta, such as hadron masses or the HVP contribution to the muon $g{-}2$, it doesn't apply to momentum-dependent observables like leptonic decay rates. In this case, the breaking of rotational symmetry in a finite volume makes the cancellation of the terms at $\Ocal(1/L^3)$ more intricate. Two methods are proposed to implement the infrared improvement in a lattice calculation. Both methods are currently under study to test numerically the differences between $\QEDL$ and $\QEDr$ prescriptions and the results will be discussed in forthcoming publications. Given that implementing the $\QEDr$ action is a straightforward modification of the $\QEDL$ prescription, we anticipate this new method could significantly contribute to reducing systematic uncertainties in various calculations of IB corrections.

\acknowledgments

I am grateful to all my collaborators, in particular M.~T.~Hansen, N.~Hermansson-Truedsson and A.~Portelli, whose contributions significantly shaped the content of this paper, as well as F.~Erben and F.~Joswig for the precious collaboration on setting up the numerical investigation of $\QEDr$. I also wish to thank A.~Patella and N.~Tantalo for the useful discussions ahead of the conference. 
I have been supported in part by the UK STFC grant ST/P000630/1. This project has also received funding from the European Research Council~(ERC) under the European Union's Horizon 2020 research and innovation programme under grant agreements No 757646 and No 813942, and from the European Union's Horizon Europe research and innovation programme under the Marie Sklodowska-Curie grant agreement No 101108006.

\bibliography{refs.bib}

\end{document}